\documentclass[aps,prl,10pt,twocolumn,superscriptaddress,showpacs,showkeys,preprintnumbers]{revtex4-2}

\usepackage[utf8]{inputenc}
\usepackage[T1]{fontenc}
\usepackage[english]{babel}
\babelprovide[import=en]{en}
\usepackage{graphicx}
\usepackage{siunitx}
\usepackage{amsmath,amssymb}
\makeatletter
\def\label#1{\@bsphack\begingroup\UseHookWithArguments{label}{1}{#1}%
\protected@write\@auxout{}%
{\string\newlabel{#1}{{\@currentlabel}{\thepage}{\@currentlabelname}{\@currentHref}{\@kernel@reserved@label@data}}}%
\endgroup\@esphack}%
\makeatother
\usepackage[colorlinks,linkcolor=blue,citecolor=blue,urlcolor=blue]{hyperref}

\begin{document}

\title{Breakdown of spallation in multi-pulse ultrafast laser ablation}

\author{David Redka}
\email{dredka@hm.edu}
\affiliation{Laser Center HM, Hochschule M\"unchen University of Applied Sciences, Germany}

\author{Julian Vollmann}
\affiliation{Laser Center HM, Hochschule M\"unchen University of Applied Sciences, Germany}

\author{Nicolas Thomae}
\affiliation{Laser Center HM, Hochschule M\"unchen University of Applied Sciences, Germany}

\author{Maximilian Spellauge}
\affiliation{Laser Center HM, Hochschule M\"unchen University of Applied Sciences, Germany}

\author{Heinz P. Huber}
\affiliation{Laser Center HM, Hochschule M\"unchen University of Applied Sciences, Germany}

\date{\today}

\begin{abstract}
    Ultrashort-pulse laser ablation of metals near damage threshold is governed by homogeneous spallation, in which tensile unloading releases a nanometre-thin liquid film whose optical signatures are temporally evolving concentric Newton rings in pump--probe experiments. This well-established picture rests almost exclusively on single-pulse results obtained on ideally flat surfaces, yet application-oriented processing invariably operates in a multi-pulse regime in which each pulse irradiates a surface progressively modified by preceding pulses. Whether homogeneous spallation persists under these conditions has remained an open question. Here we resolve this question using time-resolved pump--probe interferometry applied pulse by pulse to austenitic stainless steel. We show that homogeneous spallation dominates the first pulse, while its contribution is strongly reduced for the second pulse. By the third pulse, Newton rings vanish and sustained surface bulging collapses, with the optical transients fully saturating into a phase-explosion-like signature by the fourth pulse. Fourier-domain coherence analysis rules out roughness-induced decoherence as an optical artefact. Four independent observables, spanning time-resolved and final-state measurements, converge on the same transition after three to four pulses. Spallation-layer formation, widely invoked to explain ultrashort-pulse ablation of metals, is thus a single-pulse phenomenon rather than a multi-pulse ablation mechanism.
\end{abstract}

\maketitle

Laser material processing accounts for approximately \SI{40}{\percent} of the global laser market, with ultrashort-pulse machining alone representing a billion-dollar segment \cite{Optech2026}. In applications, material removal and structuring rely on multi-pulse irradiation \cite{sugioka_ultrafast_2014, gattass_femtosecond_2008, kerse_ablation-cooled_2016}, where each pulse interacts with the morphology left by the preceding pulses. Yet the mechanistic picture derives almost entirely from single-pulse experiments \cite{sokolowski-tinten_transient_1998, winter_ultrafast_2020, fuentesedfuf_ultrafast_2022, white_superheating_2025} and simulations \cite{lorazo_short-pulse_2003, povarnitsyn_microscopic_2015, rethfeld_modelling_2017, inogamov_nanospallation_2008}.

For metals, the single-pulse regime is sufficiently well understood to enable robust quantitative agreement between simulation and experiment \cite{povarnitsyn_wide-range_2012,chen_time-resolved_2025}. In particular, energy deposition drives quasi-isochoric lattice heating under stress confinement, resulting in GPa-level compressive stresses \cite{zhigilei_atomistic_2009, paltauf_photomechanical_2003}. Subsequent tensile unloading leads to sub-surface void nucleation \cite{rethfeld_ultrafast_2002}, coalescence and ejection of a nanometre-thick liquid film \cite{zhigilei_atomistic_2009}. This so-called spallation layer was identified early on in pump--probe microscopy (PPM) through high-contrast Newton rings (NR) \cite{sokolowski-tinten_transient_1998, bonse_time-_2006} and is widely regarded as the dominant mechanism for fluences ranging from ablation threshold $F_\mathrm{thr}$ up to approximately $3$ to $5\,F_\mathrm{thr}$ \cite{redka_mechanisms_2025}. At higher fluence, lattice temperatures approach the thermodynamic critical point, triggering phase explosion that releases a vapour--droplet mixture through explosive boiling \cite{bulgakova_pulsed_2001, lorazo_short-pulse_2003, zhigilei_atomistic_2009}, observed in PPM as a strongly absorbing and scattering ablation plume \cite{chen_time-resolved_2025, lin_ultrafast_2025}.

How far this single-pulse picture carries over to multi-pulse processing remains unresolved. Under repeated irradiation, incubation lowers the ablation threshold \cite{jee_laser-induced_1988, mannion_effect_2004}, surface roughness accumulates \cite{thomae_singlepulse_2026}, and self-organised nanostructures such as laser-induced periodic surface structures or cone-like protrusions emerge \cite{bonse_laser-induced_2017, gnilitskyi_high-speed_2017, vorobyev_direct_2013}. These feedback mechanisms, driven in part by sub-wavelength electromagnetic near-field modulation \cite{rudenko_amplification_2019, bonse_maxwell_2020, thomae_singlepulse_2026}, progressively reshape the target on which each subsequent pulse acts. Whether a homogeneous spallation layer can persist under such evolving conditions, or the ablation mechanism changes, is unknown.

Here we address this question using time-resolved pump--probe interferometry (PPI) as a function of pulse number $N$ in austenitic stainless steel AISI~304, under conditions designed to avoid heat accumulation. Beyond its industrial relevance, AISI~304 provides an unusually favourable model system because single-pulse spallation leaves low residual roughness while showing near-unity NR contrast \cite{fuentesedfuf_ultrafast_2022, redka_mechanisms_2025}. By quantifying spatiotemporal evolution of the reflectance change $\Delta R/R_0$ and interferometric phase $\Delta\phi$, supported by Fourier-domain coherence analysis and transfer-matrix modelling, we show that homogeneous spallation is a single-pulse phenomenon on pristine surfaces that breaks down within the first few pulses of multi-pulse irradiation.

\begin{figure*}[!ht]
  \centering
  \includegraphics[width=\textwidth]{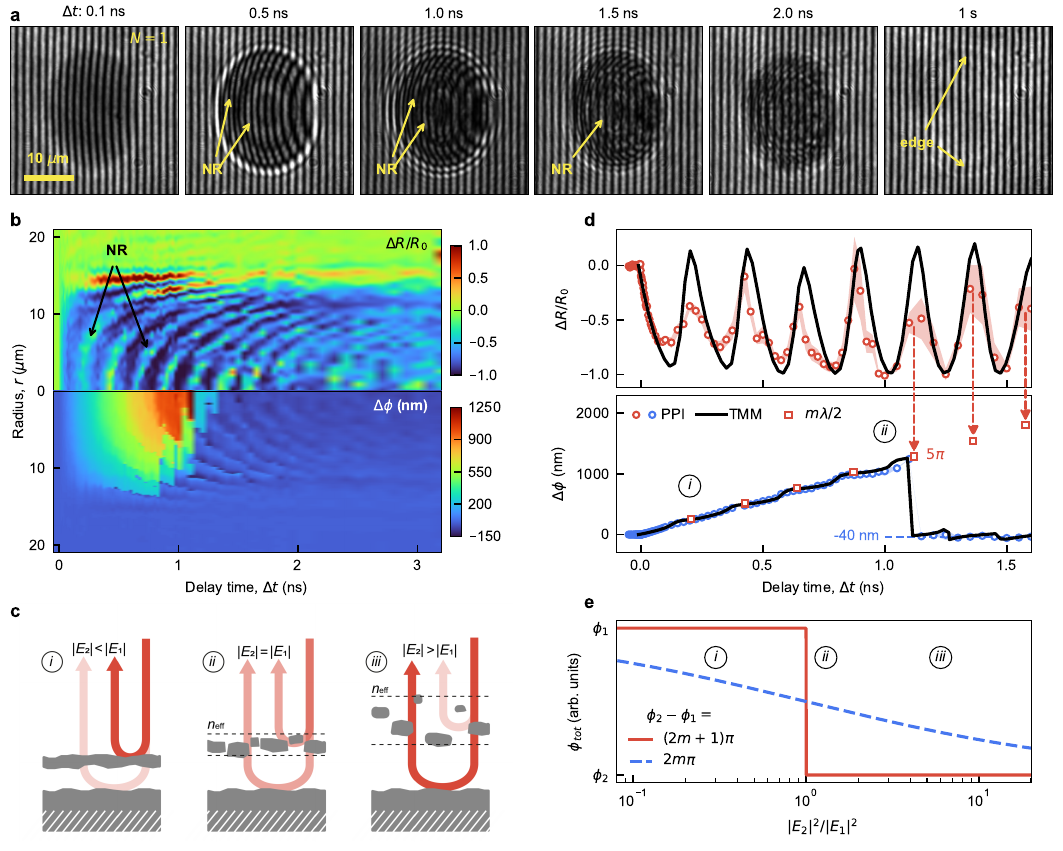}
  \caption{\textbf{Single-pulse spallation dynamics at $2F_{\mathrm{thr}}$.} (a)~Raw interferograms at selected pump--probe delays $\Delta t$ (\SI{38.7}{\micro\meter} field of view). (b)~Spatiotemporal maps of the relative reflectivity change $\Delta R/R_0$ (top) and interferometric phase $\Delta\phi$ expressed as equivalent optical path length (bottom) versus delay $\Delta t$ and radial distance $r$ from the irradiation centre. (c)~Three-stage spallation-layer disintegration pathway and the associated two-beam interference model, with field $E_1$ reflected at the spallation-layer front and $E_2$ at the remaining bulk. (d)~Centre traces of $\Delta R/R_0$ (top) and $\Delta\phi$ (bottom) compared with transfer-matrix simulations (black solid line). Red squares mark layer heights extracted from successive interference maxima via $h = m\lambda_{\mathrm{probe}}/2$. (e)~Total phase $\phi_{\mathrm{tot}}$ as a function of $|E_2|^2/|E_1|^2$ for selected phase differences $\phi_2 - \phi_1$ in the two-beam model.}
  \label{fig:f1}
\end{figure*}

\section*{Single-pulse spallation}

Single-pulse PPI employed a \SI{300}{\femto\second}, \SI{1030}{\nano\meter} pump at a peak fluence of $2F_{\mathrm{thr}}$ and a \SI{250}{\femto\second}, \SI{515}{\nano\meter} probe. Raw interferograms (Fig.~1a) show straight spatial-carrier fringes (SCF) for an undeformed surface, where right- and left-curved SCF indicate a decreased and increased optical path, respectively~\cite{temnov_ultrafast_2006, inogamov_nanospallation_2008}. At $\Delta t=\SI{0.1}{\nano\second}$, intensity drops with weakly right-curved SCF, corresponding to bulging. Between $\Delta t=\SIrange{0.5}{1.5}{\nano\second}$, concentric NR evidence thin-film interference between the spallation layer and the remaining bulk~\cite{bonse_time-_2006, fuentesedfuf_ultrafast_2022}. At $\Delta t=\SI{2}{\nano\second}$, weakly left-curved SCF and missing NR indicate a signal dominated by the underlying target \cite{redka_mechanisms_2025}. The final state ($\Delta t=\SI{1}{\second}$) shows no reflectivity change, consistent with specular reflectance measurements \cite{thomae_deciphering_2026}, and an SCF kink at the crater edge matches the rectangular crater profile \cite{ionin_material_2017}.

$\Delta R/R_0$ and $\Delta\phi$ were obtained from Fourier analysis (Methods). $\Delta R/R_0$ (Fig.~1b, top) quantitatively matches prior PPM on stainless steel~\cite{winter_ultrafast_2020, fuentesedfuf_ultrafast_2022, redka_mechanisms_2025}. Near-unity NR contrast indicates minimal absorption, characteristic of pure spallation-driven thin-film interference \cite{fuentesedfuf_ultrafast_2022}. NR encode radial parabolic bulging and spallation-layer velocities set by the Gaussian fluence profile~\cite{lin_ultrafast_2025, fuentesedfuf_ultrafast_2022, inogamov_nanospallation_2008, wu_microscopic_2014} and remain visible up to \SI{2}{\nano\second} in the crater centre. Using $h=m\lambda_{\mathrm{probe}}/2$ and a vacuum-like interlayer for $\Delta t>\SI{200}{ps}$~\cite{chen_time-resolved_2025, fuentesedfuf_ultrafast_2022}, successive maxima at $r=0$ yield a constant layer velocity of \SI{1.11(1)}{\kilo\meter\per\second}.

The phase is expressed as an optical path length in units of nm via $\Delta\phi = -\Delta\phi_\mathrm{rad}\lambda_\mathrm{probe}/4\pi$, with positive $\Delta\phi$ indicating bulging. This relation is exact only for a purely geometric phase, as transient refractive-index changes of the material add a $\Delta\phi$ \cite{temnov_ultrafast_2006}. Pump-probe ellipsometry limits this contribution to $\SI{-12}{\nano\meter}$ (Supplement) \cite{winter_ultrafast_2020}, a secondary contribution when exceeding the first hundreds ps, due to layer velocities of \SI{1}{\nano\meter\per\pico\second}. Figure~1d (top) shows $\Delta\phi$ rising monotonically from zero delay, reproducing the same constant velocity of \SI{1.11(1)}{\kilo\meter\per\second}. Whereas $\Delta R/R_0$ tracks the spallation layer up to \SI{2}{\nano\second}, phase tracking remains robust only up to $\SI{1.1}{\nano\second}$, corresponding to a bulging height of \SI{1.25}{\micro\meter} ($\approx 2.5\lambda_{\mathrm{probe}}$, Fig.~1d, bottom). We attribute this discrepancy to transient optical thinning of the spallation layer. Beyond \SI{1.1}{\nano\second}, $\Delta\phi$ drops to $\SI{-40}{\nano\meter}$ and remains constant. In the final state ($\Delta t=\SI{1}{\second}$), $\Delta\phi=\SI{-14(1)}{\nano\meter}$ reflects the crater depth and agrees with confocal microscopy (Supplement).

To account for optical thinning we propose the three-stage pathway in Fig.~1c, supported by transfer-matrix simulations (Methods). With a parabolic velocity profile and an initial layer thickness of \SI{14}{\nano\meter}, optical thinning is captured, to first order, by a linear decrease of the steel volume fraction and $n_{\mathrm{eff}}(\Delta t)$ under mass conservation, using a Bruggeman effective-medium approximation \cite{niklasson_effective_1981}. The resulting centre traces reproduce the measured $\Delta R/R_0$ and $\Delta\phi$ (Fig.~1d).

Stage~(\textit{i}), $\Delta t\leq\SI{0.1}{\nano\second}$, covers layer detachment and initial constant-velocity propagation. Experiment and simulation both exhibit an initial $\Delta R/R_0$ minimum at $\Delta t\approx\SI{0.11}{\nano\second}$, set by destructive thin-film interference and thus by layer thickness, velocity and interlayer absorptance~\cite{redka_mechanisms_2025}. The deeper simulated minimum ($\Delta R/R_0\approx-0.9$ versus $\approx-0.7$ experiment) reflects an initially evolving spongy interlayer of interconnected liquid bridges~\cite{zhigilei_atomistic_2009} whose refractive index relaxes towards vacuum on a \SI{100}{\pico\second} timescale~\cite{chen_time-resolved_2025, fuentesedfuf_ultrafast_2022}.

Stage~(\textit{ii}), $\SI{0.1}{\nano\second}<\Delta t\leq\SI{2}{\nano\second}$, comprises continued spallation layer propagation with progressive destabilisation, partial fracture and fragmentation, resulting in optical thinning. Thin liquid films of nanometre thickness are inherently rupture-prone owing to thermal fluctuations and intrinsic inhomogeneities~\cite{rahman_life_2024, duran-olivencia_instability_2019}. During propagation these perturbations amplify, further assisted by hydrodynamic instabilities \cite{shih_generation_2017}, driving lateral perforation and subsequent hole coalescence~\cite{rahman_life_2024}. On real surfaces, electromagnetic near-field scattering additionally imprints wavelength-scale modulations of the absorbed fluence \cite{rudenko_light_2019, rudenko_high-frequency_2020}. Higher local fluences result in faster spallation velocities but thinner layers~\cite{wu_microscopic_2014}, so that lateral fluence heterogeneity translates into axial segmentation, complementing lateral perforation. The initially homogeneous spallation layer thus progressively breaks up, observed experimentally as optical thinning. In this stage, $\Delta\phi$ increases with small oscillations (aligned with $\Delta R/R_0$ minima) up to \SI{1.1}{\nano\second}, followed by an abrupt drop, whereas NR persist throughout (Fig.~1d). A two-beam interference model captures this behaviour. The fields $E_1$ reflected at the spallation-layer front and $E_2$ at the bulk give $\phi_{\mathrm{tot}}=\arg\left(E_1\mathrm{e}^{\mathrm{i}\phi_1}+E_2\mathrm{e}^{\mathrm{i}\phi_2}\right)$. Optical thinning increases transmission and thus $|E_2|^2/|E_1|^2$, shifting $\phi_{\mathrm{tot}}$ from $\phi_1$ when $|E_2|^2/|E_1|^2<1$ towards $\phi_2$ when $|E_2|^2/|E_1|^2>1$, with equal weighting at unity (Fig.~1e). The transition sharpness depends on $\phi_2-\phi_1$. For odd multiples of $\pi$ the fields are antiparallel and the phase jump is discontinuous, whereas for even multiples the fields are parallel and the phase remains insensitive to the amplitude ratio. As $\phi_2-\phi_1$ evolves, the system cycles between these limits with growing oscillation amplitude. The observed phase drop coincides with the fifth maximum in $\Delta R/R_0$, at $\phi_2-\phi_1\approx5\pi$ ($2.5\lambda_{\mathrm{probe}}$). The interferometric phase is therefore more sensitive to optical thinning than NR contrast, providing an early indicator of spallation-layer fragmentation.

Stage~(\textit{iii}) sets in once the film fully fragments, eliminating the optical interface, consistent with the disappearance of NR beyond $\SI{2}{\nano\second}$. In the transfer-matrix model, for $\Delta t>\SI{1.1}{\nano\second}$ the phase approaches $\SI{-30}{\nano\meter}$ (experiment $\SI{-40}{\nano\meter}$), exceeding the final crater depth of $\SI{-14}{\nano\meter}$. The excess optical path arises from residual ejecta remaining in the probe path, forming a dilute medium with $n_{\mathrm{eff}}>1$.

\begin{figure*}[!ht]
  \centering
  \includegraphics[width=\textwidth]{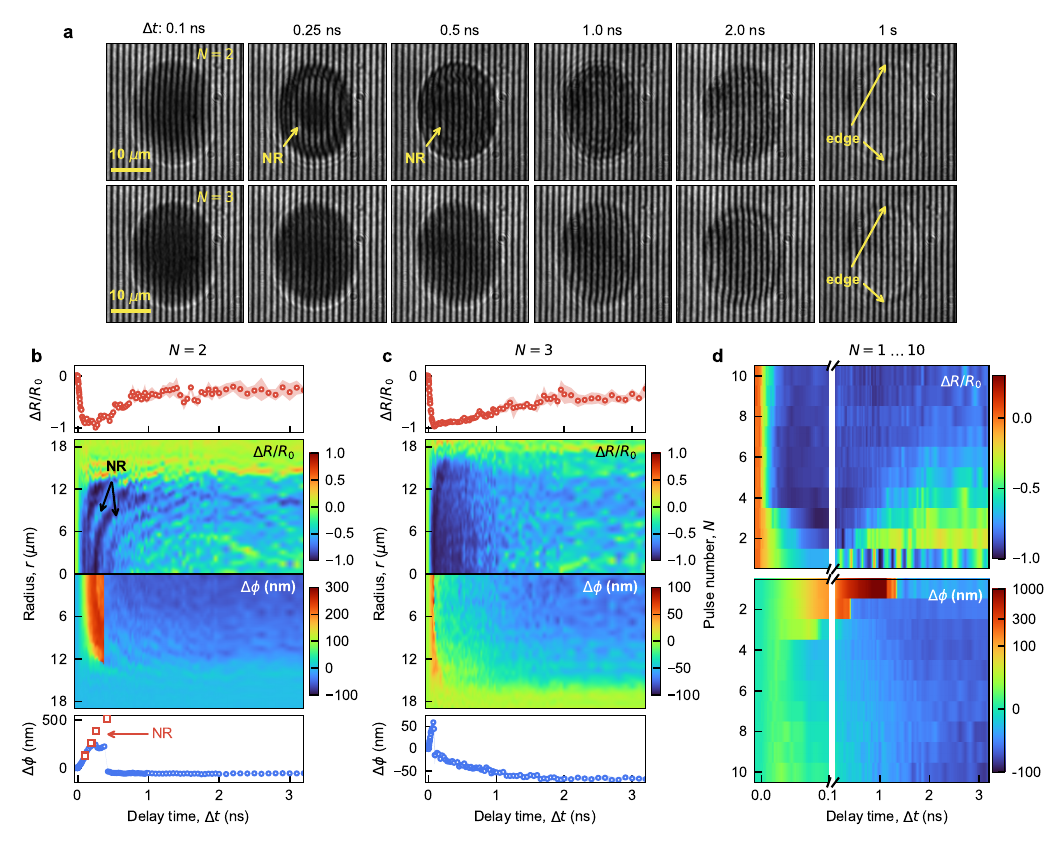}
  \caption{\textbf{Breakdown of spallation under multi-pulse irradiation at $2F_{\mathrm{thr}}$.} (a)~Raw interferograms after $N=2$ (top) and $N=3$ (bottom) pulses per site, with $\Delta t$ referenced to the respective pump pulse (\SI{38.7}{\micro\meter} field of view). (b)~Spatiotemporal maps of $\Delta R/R_0$ (top) and equivalent optical path length change $\Delta\phi$ (bottom) for $N=2$, referenced to the final state after the preceding pulse, with centre traces shown next to each map. (c)~As in (b), for $N=3$. (d)~Centre transients for $N=1$ to $10$ compiled as a function of pulse number and delay, with colour encoding $\Delta R/R_0$ (top) and $\Delta\phi$ (bottom).}
  \label{fig:f2}
\end{figure*}

\section*{Multi-pulse ablation}
PPI interferograms were recorded under otherwise identical conditions with up to $N=10$ pulses per position and an inter-pulse delay of \SI{1}{\second} to suppress heat accumulation \cite{weber_heat_2014, bauer_heat_2015}. Incubation nevertheless lowers the ablation threshold with increasing pulse number according to $F_{\mathrm{thr}}(N)=F_{\mathrm{thr}}(1)\,N^{S-1}$ \cite{jee_laser-induced_1988}. From $D^2$-measurements \cite{liu_simple_1982}, we obtain $S=\num{0.74(1)}$ (Supplement), consistent with values reported for stainless steel \cite{di_niso_role_2014, lickschat_fundamental_2020}. The effective normalised fluence $F/F_{\mathrm{thr}}(N)$ thus increases to approximately $3$ at $N=5$ and $3.7$ at $N=10$, nominally within the single-pulse spallation regime \cite{redka_mechanisms_2025}.

Raw interferograms for $N=2$ and $N=3$, with $\Delta t$ referenced to the $N$-th pulse, are shown in Fig.~2a. For $N=2$, the early response partially resembles $N=1$. Reduced intensity with right-curved SCF at $\Delta t=\SI{0.1}{\nano\second}$ indicates bulging, and NR appear at $r=\SI{8}{\micro\meter}$ for $\Delta t$ between $\SI{0.25}{\nano\second}$ and $\SI{0.5}{\nano\second}$. At larger radii SCF remain right-curved, whereas the centre becomes weakly left-curved, indicating increased sensitivity to the underlying bulk. For $\Delta t>\SI{1.5}{\nano\second}$, the spot appears dim with weakly left-curved SCF. In the final state, SCF are straight in the crater centre with a distinct kink at the edge and no overall reflectivity decrease.

Spatiotemporal maps of $\Delta R/R_0$ and $\Delta\phi$ for $N=2$ are displayed in Fig.~2b together with centre traces, using the final state after the $(N-1)$-th pulse as reference to isolate the per-pulse response (Methods). $\Delta R/R_0$ exhibits NR up to $\Delta t\approx\SI{1}{\nano\second}$ for $\SI{6}{\micro\meter}<r<\SI{12}{\micro\meter}$, whereas the central region shows strongly reduced signal with nearly vanishing NR contrast. NR analysis yields a surface velocity of $\SI{1.3(2)}{\kilo\meter\per\second}$ and a maximum spallation-layer height of \SI{515}{\nano\meter}. For $\Delta t>\SI{0.5}{\nano\second}$, reflectivity recovers rapidly and stabilises at $\Delta R/R_0\approx -0.3$ by $\Delta t\approx\SI{1}{\nano\second}$. The $\Delta\phi$ map shows the same parabolic bulging to $N=1$ up to \SI{0.4}{\nano\second}, but with a strongly altered crater-centre response. At $r=0$, $\Delta\phi$ reaches \SI{243}{\nano\meter} at $\Delta t=\SI{0.25}{\nano\second}$, corresponding to a velocity of $\SI{1.27(5)}{\kilo\meter\per\second}$, then jumps towards zero at $\Delta t=\SI{0.4}{\nano\second}$ and relaxes to approximately \SI{-60}{\nano\meter} by $\Delta t=\SI{0.5}{\nano\second}$. As for $N=1$, this exceeds the final crater depth of \SI{-16}{\nano\meter}.

The dynamics change qualitatively at $N=3$. NR are absent in the raw interferograms (Fig.~2a), and SCF no longer show initial right-curvature. Instead, at $\Delta t=\SI{0.1}{\nano\second}$, intensity drops strongly and the central region ($r<\SI{6}{\micro\meter}$) exhibits pronounced loss of SCF contrast, indicating strong attenuation and scattering of the probe intensity. For $\Delta t>\SI{0.25}{\nano\second}$, SCF contrast recovers with slight left-curvature while overall intensity remains reduced. As for all preceding pulse numbers, no final-state reflectivity change is observed inside the crater with straight, high-intensity SCF, in agreement with specular-reflectance measurements \cite{thomae_deciphering_2026}, confirming that the $(N-1)$-th final state serves as a reliable reference for isolating the $N$-th pulse response.

In the $\Delta R/R_0$ map for $N=3$ (Fig.~2c), NR are completely absent. $\Delta R/R_0$ drops to nearly $-1$ at $\Delta t=\SI{0.1}{\nano\second}$ and recovers slowly towards $\approx -0.5$ by approximately \SI{2}{\nano\second}. This transient resembles the strongly absorbing and scattering response reported for single-pulse phase explosion above $5\,F_{\mathrm{thr}}$ \cite{chen_time-resolved_2025, redka_mechanisms_2025, lin_ultrafast_2025}, although the incubation-law estimate of approximately $2.7\,F_{\mathrm{thr}}$ for $N=3$ would place the process in the nominal single-pulse spallation regime \cite{redka_mechanisms_2025}. The corresponding $\Delta\phi$ (Fig.~2c, bottom) shows only weak initial bulging. At $r=0$, $\Delta\phi$ reaches approximately \SI{50}{\nano\meter} at $\Delta t=\SI{70}{\pico\second}$, corresponding to a centre velocity of $\SI{0.9(1)}{\kilo\meter\per\second}$, then drops abruptly to $\approx\SI{-15}{\nano\meter}$ at $\Delta t=\SI{0.1}{\nano\second}$ and decreases continuously to $\approx\SI{-67}{\nano\meter}$ at $\Delta t=\SI{2}{\nano\second}$. By contrast, the final crater depth for $N=3$ is \SI{-23}{\nano\meter}.

For $N>3$, the qualitative behaviour remains unchanged. Corresponding maps up to $N=10$ are provided in the Supplement. Fig.~2d compiles centre transients as functions of $N$. At $N=4$, the onset of the reflectivity drop to $\Delta R/R_0\approx -1$ shifts from $\Delta t\approx\SI{110}{\pico\second}$ to $\approx\SI{40}{\pico\second}$ and remains essentially constant for $N>4$, whereas the subsequent recovery sets in at progressively later delays with increasing $N$. The $\Delta\phi$ response at $N=4$ shows reduced initial bulging, reaching approximately \SI{25}{\nano\meter} at $\Delta t\approx\SI{20}{\pico\second}$, and remains similar for $N>4$. Beyond \SI{20}{\pico\second}, phase decreases gradually and saturates near \SI{-70}{\nano\meter} for $\Delta t\geq\SI{2.5}{\nano\second}$, while the final crater-depth increment remains around \SI{-20}{\nano\meter}. Taken together, the evolution from $N=1$ to $N\geq 3$ marks a transition from homogeneous spallation, characterised by high-contrast NR and sustained surface bulging, to phase-explosion-like optical dynamics in which strong scattering and absorption suppress the reflected signal and no homogeneous spallation layer is observed.

\begin{figure*}[!ht]
  \centering
  \includegraphics[width=\textwidth]{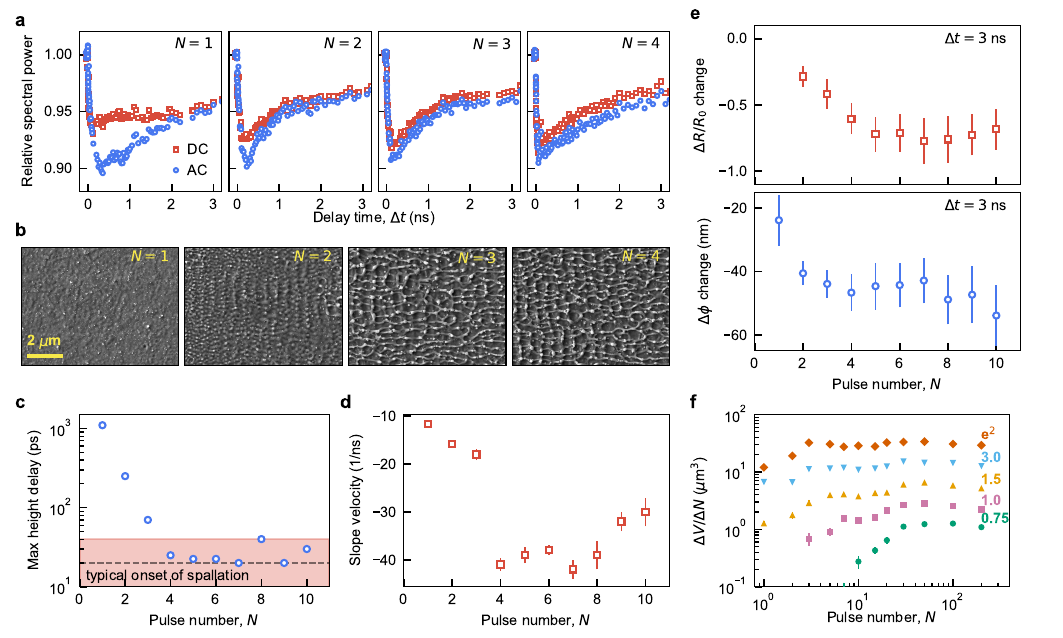}
  \caption{\textbf{Coherence diagnostics and pulse-resolved ablation metrics establish a spallation-to-phase-explosion like transition after three pulses.} (a)~Normalised spectral powers of the DC (red) and AC (blue) Fourier components as a function of delay for $N=1$, $2$, $3$, and $4$, extracted within a restricted integration bandwidth (Methods). (b)~Scanning electron micrographs of the crater-centre morphology (\SI{10}{\micro\meter} field of view) after the $N$-th pulse at $3F_{\mathrm{thr}}$. (c)~Delay at which surface tracking from the spallation-layer front breaks down versus pulse number. Maximum heights are annotated and the red shaded band marks the typically predicted onset of spallation of \SI{10}{ps} to \SI{40}{ps}. (d)~Initial slope $\Delta(\Delta R/R_0)/\Delta t$ up to the first reflectivity minimum. (e)~$\Delta R/R_0$ at $\Delta t = \SI{3}{\nano\second}$, corrected for the final-state reflectance to isolate the transient contribution. (f)~$\Delta\phi$ at $\Delta t = \SI{3}{\nano\second}$, corrected for the final crater depth. (g)~Ablated volume per pulse $\Delta V/\Delta N$ versus pulse number $N$, measured for fluences between $0.75\,F_{\mathrm{thr}}$ and $e^2 F_{\mathrm{thr}}$ over $N = 1$ to $200$.}
  \label{fig:f3}
\end{figure*}

\section*{Sustained spatial coherence}
The disappearance of NR for $N\geq 3$ could, in principle, arise from roughness-induced loss of spatial coherence rather than a change in ablation mechanism. We use the interferometric data as a direct, time-resolved diagnostic to rule out this possibility. In Fourier space, the raw interferograms exhibit separable DC and AC spectral peaks \cite{takeda_fourier-transform_1982}, with an integration bandwidth set by the reciprocal beam waist $1/w_0$. For a balanced interferometer at negative delay ($I_{\mathrm{ref}}\approx I_{\mathrm{obj}}$), $I_{\mathrm{DC}}=I_{\mathrm{ref}}+I_{\mathrm{obj}}$ and $I_{\mathrm{AC}}=2\sqrt{I_{\mathrm{ref}}I_{\mathrm{obj}}}\,\gamma$ \cite{torcal-milla_modified_2023}, with $\gamma$ the degree of coherence, so that the AC/DC ratio serves as a practical diagnostic of coherence loss. Crucially, within a finite integration bandwidth the ratio is also sensitive to the spatially structured phase. A bulging spallation layer redistributes AC spectral weight beyond the integration window, reducing AC/DC even for constant $\gamma$ (Supplement). By contrast, spatially smooth intensity changes such as homogeneous absorption broaden both spectral peaks identically and therefore preserve AC/DC ratio. Diffuse scattering from either the spallation layer or the bulk constitutes an incoherent background within the DC signal without providing contributions to the AC term, causing any roughness-induced decoherence to manifest as a drop in the AC/DC ratio.

Normalised DC and AC spectral powers are shown in Fig.~3a for pulse numbers up to $N=4$. For $N=1$ and $N=2$, AC power decreases strongly relative to DC up to delays of \SI{2}{\nano\second} for $N=1$ and approximately \SI{0.5}{\nano\second} for $N=2$. This mismatch coincides with the time window over which NR are observed (Fig.~1a, Fig.~2b), consistent with spectral-weight redistribution of the AC peak by the structured phase of the bulging spallation layer. At late delays, once $\Delta R/R_0$ has recovered and no structured phase remains, AC and DC converge, confirming $\Delta\gamma\approx 0$ for the post-ablation surface. The duration of the AC/DC mismatch therefore serves as a practical proxy for spallation-layer lifetime.

For $N\geq 3$, no systematic AC/DC mismatch is detected beyond percent-level differences, despite the strong reflectivity drop in the crater centre ($\Delta R/R_0\approx -1$). Both spectral peaks decrease proportionally within the Fourier-space integration window, preserving AC/DC $\approx 1$. The absence of mismatch confirms that the probe remains spatially coherent throughout and that the disappearance of NR reflects a genuine absence of a homogeneous spallation layer rather than an optical artefact.

Surface roughness measurements independently support this conclusion. Scanning electron microscopy images at the crater centre are shown in Fig.~3b. The RMS roughness $S_q$, obtained from crater-centre topographies via confocal microscopy, increased from \SI{6}{\nano\meter} after the first pulse to \SI{9}{\nano\meter} ($N=2$) and \SI{20}{\nano\meter} ($N=3$) at $3\,F_{\mathrm{thr}}$, reaching \SI{50}{\nano\meter} at $N=10$. At $1.5\,F_{\mathrm{thr}}$, $S_q$ remained below \SI{10}{\nano\meter} for $N<4$. Within scalar scattering theory \cite{bennett_relation_1961}, $R_\mathrm{s}/R_0 = \exp(-(4\pi S_q/\lambda)^2)$ yields $R_\mathrm{s}\approx 0.95$ for $S_q<\SI{10}{\nano\meter}$ and $\approx 0.80$ at $S_q=\SI{20}{\nano\meter}$. In the two-beam picture of Fig.~1e, roughness on either reflecting surface diminishes its specular field amplitude, making $|E_1|^^2$ and $|E_2|^2$ increasingly unequal. This reduces NR contrast while $\phi_{\mathrm{tot}}$ shifts smoothly towards the stronger field. Thus, the $N=1$ data illustrate optical thinning and not transient roughness. NR persist to \SI{2}{\nano\second} while phase tracking fails already at \SI{1.1}{\nano\second} once $|E_2|^2$ exceeds $|E_1|^2$. The abrupt and simultaneous disappearance of both NR and bulging at $N=3$ is therefore incompatible with gradual roughness-driven degradation and confirms again the absence of a homogeneous spallation layer.

\section*{Discussion}
The delay of maximum bulging height (Fig.~3c) quantifies the pulse-to-pulse evolution. From $N=1$ to $N=3$, the characteristic time decreases exponentially from $\SI{1.1}{\nano\second}$ to $\SI{70}{\pico\second}$ and saturates at approximately $\SI{20}{\pico\second}$ for $N>3$, with bulging limited to $\SI{25}{\nano\meter}$. As previously noted, this apparent displacement remains convolved with an optical phase shift arising from transient refractive-index changes, precluding a purely geometric interpretation. Atomistic and hydrodynamic simulations place tensile-stress-driven void nucleation at $\SIrange{10}{40}{\pico\second}$ and complete material separation from $\SI{50}{\pico\second}$ onward  \cite{iabbaden_atomistic_2026, wu_generation_2015, amouye_foumani_atomistic_2021, povarnitsyn_hydrodynamic_2014, abou-saleh_spallation-induced_2018, zhigilei_atomistic_2009}. Delays below $\SI{20}{\pico\second}$ therefore correspond to thermoelastic expansion and stress unloading \cite{shugaev_thermodynamic_2019} rather than a detached spallation layer. For $N>3$, the tracked optical phase accordingly reflects an evolving, phase-explosion-like state without a well-defined Fresnel interface.

The progressive destabilisation of the single-pulse spallation mechanism is most consistently explained by accumulated morphological modification of the target surface. Pulse-to-pulse induced inhomogeneities in layer thickness and velocity, seeded by surface roughness \cite{sun_dynamics_2025}, subsurface voids \cite{shugaev_thermodynamic_2019} and sub-wavelength near-field enhancement \cite{rudenko_light_2019, rudenko_amplification_2019, rudenko_high-frequency_2020, tsibidis_synergy_2022, romer_finite-difference_2014}, accelerate spallation-layer fragmentation (as described in stage~ii of Fig.~1c) through both lateral perforation and axial segmentation. Finite-difference time-domain simulations on real crater topographies quantify the field enhancement \cite{thomae_singlepulse_2026}, showing that surface roughness of $S_q=\SI{6}{\nano\meter}$, $\SI{9}{\nano\meter}$ and $\SI{20}{\nano\meter}$ produces local peak-to-valley fluence variations of approximately $30\%$, $50\%$ and $120\%$ at sub-$\si{\micro\meter}$ lateral scale, while total absorbed energy remains unchanged \cite{thomae_deciphering_2026}. With increasing $N$ the resulting near-field enhancement is expected to trigger earlier layer collapse and spatially localised phase-explosion-like ablation, so that at higher pulse numbers no homogeneous spallation layer forms.

The initial slope of $\Delta R/R_0$ (Fig.~3d) corroborates this picture. The slope steepens progressively from $N=1$ to $N=3$, strongly drops for $N=4$ and subsequently saturates, whereas the initial bulging velocity at $\Delta t<\SI{20}{\pico\second}$ remains nearly constant at approximately $\SI{1}{\kilo\meter\per\second}$ over all $N$ (Supplement). The accelerated reflectivity decay is therefore consistent with enhanced absorption and scattering in the transient ablation plume, mirroring the single-pulse spallation-to-phase-explosion transition with increasing fluence \cite{chen_time-resolved_2025, redka_mechanisms_2025, lin_ultrafast_2025}. At longer delays (Fig.~3e), $\Delta R/R_0$ at $\Delta t=\SI{3}{\nano\second}$, corrected for the final-state reflectance, decreases monotonically up to $N=7$ and saturates at approximately $-0.8$. The corresponding corrected $\Delta\phi$ (Fig.~3f) saturates at $\approx\SI{-45}{\nano\meter}$ by the third pulse. The concurrent growth of absorption and transient optical path length beyond crater depth is consistent with increasingly dense or extended ablated material above the surface.

The transition is equally evident in ablation energetics. The ablated volume per pulse $\Delta V/\Delta N$ (Fig.~3g), measured for fluences between $0.75\,F_{\mathrm{thr}}$ and $\mathrm{e}^2\,F_{\mathrm{thr}}$ (single-pulse thresholds) over $N=1$ to $N=200$, increases over the first three pulses and then converges to a stable plateau. Since both self-reflectance and FDTD-computed total absorption  \cite{thomae_deciphering_2026} remain quasi-constant over the early pulse range ($N<10$), this evolution reflects changes in ablation dynamics rather than in total absorbed energy. The qualitative transition at $N=3$ and the subsequent dynamical saturation at $N=4$ together define a crossover window that coincides with saturation of surface bulging (Fig.~3c), stabilisation of the initial $\Delta R/R_0$ slope (Fig.~3d) and saturation of the corrected transient phase (Fig.~3f), providing four independent observables that converge on the same pulse number.

Our observations indicate that multi-pulse irradiation drives the surface beyond a degree of morphological modification at which homogeneous spallation can no longer be sustained. Material removal instead proceeds through predominantly phase-explosion-like optical dynamics, even when incubation-law estimates place excitation within the nominal single-pulse spallation regime. Spallation-layer formation, widely invoked to explain ultrashort-pulse ablation of metals, thus emerges here as a single-pulse mechanism on pristine surfaces rather than a multi-pulse ablation mechanism. A key driver consistent with our observations is sub-wavelength electromagnetic near-field modulation of absorbed fluence on progressively roughened surfaces, long recognised as essential for final surface morphology including laser-induced periodic surface structure formation \cite{bonse_laser-induced_2017, thomae_singlepulse_2026, rudenko_high-frequency_2020, tsibidis_synergy_2022} but less discussed in the broader context of ablation mechanism. Our results show that such first-order effects must be incorporated for a correct understanding of morphology-mediated multi-pulse processing and that single-pulse dynamics cannot simply be extrapolated to this regime.

\begin{acknowledgments}
This work was supported by the Deutsche Forschungsgemeinschaft under Grant No.\ 528706678.
H.P.H and D.R.\ want to thank Leonid V.\ Zhigilei and Evgeny A.\ Gurevich for fruitful discussions during the preparation of the manuscript.
\end{acknowledgments}

\section*{Methods}

\subsection*{Ablation target}
Austenitic stainless steel AISI~304 discs (diameter \SI{30}{\milli\meter}, thickness \SI{500}{\micro\meter}) were mirror-polished through successive grinding (SiC, down to \SI{15}{\micro\meter} grit), diamond polishing (\SIlist{9;3;1}{\micro\meter}) and colloidal $\mathrm{SiO_2}$ finishing (\SI{0.05}{\micro\meter}), yielding a root-mean-square roughness $S_q < \SI{3}{\nano\meter}$, determined via confocal microscopy.

\subsection*{Pump--probe interferometry}
A femtosecond laser (PHAROS, Light Conversion) delivered \SI{300}{\femto\second} pulses at $\lambda = \SI{1030}{\nano\meter}$ (bandwidth \SI{5.7}{\nano\meter}), \SI{400}{\micro\joule} pulse energy and \SI{500}{\hertz} repetition rate. A polarising beam splitter~(PBS) divided the beam into pump and probe with a $100{:}1$ energy ratio. Pump energy was controlled via a half-wave plate~(HWP) and PBS. A mechanical shutter selected single pulses at \SI{1}{\hertz} to suppress heat accumulation~\cite{weber_heat_2014}. The pump was focused at \SI{38.7}{\degree} incidence ($p$~polarisation) with an $f = \SI{100}{\milli\meter}$ lens to a spot size $w_0 = \SI{17(1)}{\micro\meter}$ ($1/e^2$, $M^2 = \num{1.02}$, $z_R = \SI{0.9(1)}{\milli\meter}$), yielding an ablation threshold $F_{\mathrm{thr}} = \SI{0.29(3)}{\joule\per\centi\meter\squared}$ via the $D^2$-method~\cite{liu_simple_1982} (Supplement).

The probe was frequency-doubled to \SI{515}{\nano\meter} in BBO and delayed over $\Delta t = \SI{-50}{\pico\second}$ to \SI{3.5}{\nano\second} via a motorised stage. In a Michelson-type interferometer~\cite{temnov_ultrafast_2006, pflug_ultrafast_2024}, a PBS split the probe into sample and reference arms. In the sample arm, circularly polarised light (quarter-wave plate) was focused through a $50\times$ Mitutoyo (NA = 0.42) long working distance objective onto the target at normal incidence. The reference arm used an identical layout with a tilted silver mirror.

Back-reflected beams were imaged onto a 14-bit CCD (PCO pco.pixelfly~USB, $1392 \times 1024$~pixels; field of view $\SI{179}{\micro\meter} \times \SI{132}{\micro\meter}$; pixel size \SI{0.129}{\micro\meter}). Orthogonal polarisations from the PBS were brought to interference by a \SI{45}{\degree} analyser, with the reference wavefront tilt producing spatial-carrier fringes~(SCF) of approximately \SI{2}{\micro\meter} spacing. For each delay, data were acquired on a pristine preconditioned with the corresponding number of preceding pulses. Three interferograms were recorded per delay: before irradiation, at $\Delta t$, and after irradiation. Time zero was calibrated via the Kerr response of an ITO reference. The spatial resolution of the imaging system is approximately \SI{0.75}{\micro\metre}, set by the Rayleigh criterion ($0.61\,\lambda_\mathrm{probe}/\mathrm{NA}$), and the temporal resolution is \SI{250}{\femto\second}, given by the probe pulse duration.

The relative phase shift $\Delta\phi$ and reflectance change $\Delta R/R_0$ were evaluated with respect to the pre-irradiation interferogram. In multi-pulse measurements, the $(N{-}1)$-th final state served as reference for the $N$-th pulse. $\Delta\phi$ and $\Delta R/R_0$ were retrieved by two-dimensional Fourier-transform analysis~\cite{takeda_fourier-transform_1982}. The AC sideband was isolated by an elliptical mask in reciprocal space, shifted to the origin and inverse-transformed. The squared modulus yielded reflectance. The argument yielded phase, confined to $[-\pi,\pi]$. Phase discontinuities were removed by path-dependent spatial unwrapping~\cite{takeda_fourier-transform_1982} (Supplement).

\subsection*{Transfer-matrix method}
A one-dimensional transfer-matrix code~\cite{redka_mechanisms_2025} was extended to two dimensions by discretising the surface into a $1392 \times 1024$ array of independent normal-incidence columns (grid spacing \SI{0.129}{\micro\meter}), neglecting lateral coupling, diffraction and ray-angle effects.

Within a radius of \SI{13}{\micro\meter} (the experimentally observed spallation region), the vertical stack comprised a \SI{14}{\nano\meter} spallation layer, an interlayer and the underlying bulk. A parabolic velocity profile (peak \SI{1.11}{\kilo\meter\per\second}, edge \SI{0}{\kilo\meter\per\second}) governed the temporal geometry evolution; outside this zone, only bulk material was present.

Complex refractive indices were set to $n_{\mathrm{air}} = 1$ (interlayer) and $n_{\mathrm{steel}} = 1.5 - \mathrm{i}\,2.5$ (liquid stainless steel~\cite{winter_ultrafast_2020}) for the spallation layer and bulk. Layer disintegration was modelled by reducing the effective refractive index via the Bruggeman effective-medium approximation~\cite{niklasson_effective_1981}, with a time-dependent layer thickness enforcing mass conservation. Synthetic interferograms were generated by coherent superposition with a tilted reference plane wave (carrier frequency \SI{0.5}{\per\micro\meter}) and processed with the same algorithm as the experimental data.

\subsection*{Ablation morphology and volume}
Post-ablation characterisation used $w_0 = \SI{15(1)}{\micro\meter}$, $\tau_p = \SI{515}{\femto\second}$ ($\lambda = \SI{1040}{\nano\meter}$) at \SI{35}{\degree} incidence, giving $F_{\mathrm{thr}} = \SI{0.26(1)}{\joule\per\centi\meter\squared}$ ($D^2$-method; Supplement). Craters were produced for $N < 200$ at fluences of $0.75$, $1$, $1.5$, $3$ and $e^2\,F_{\mathrm{thr}}$, with five repetitions each. Topographies were acquired by confocal microscopy ($50\times$, $\mathrm{NA} = 0.5$) and analysed in Gwyddion for roughness and ablated volume. Crater morphology was additionally imaged by scanning electron microscopy (Tescan Lyra~3).

\bibliographystyle{apsrev4-2}
\bibliography{references}

\end{document}